\documentclass[12pt,preprint]{aastex}
\usepackage{natbib}
\usepackage{epsfig}
\usepackage{graphicx}
\usepackage{subfigure}
\usepackage{float}
\usepackage{amsmath}
\usepackage{color}
\usepackage{amssymb}
\usepackage{amsfonts}
\usepackage{units}
\usepackage{mathtools}
\usepackage{bm}
\usepackage[colorlinks,linkcolor=blue,anchorcolor=green,citecolor=blue]{hyperref}
\def\be{\begin{eqnarray}}
\def\ee{\end{eqnarray}}

\newcommand{\ud}{\mathrm{d}}

\newcommand{\second}{\text{ s}}

\newcommand{\kilometer}{\text{ km}}
\newcommand{\parsec}{\text{ pc}}
\newcommand{\kiloparsec}{\text{ kpc}}

\newcommand{\megayear}{\text{ Myr}}

\begin{document}

%\psfrag{Msun}[c][][1.75]{M$_{\odot}$}

\title{Fast radio bursts from activities of neutron stars newborn in BNS mergers: offset, birth rate and observational properties}

\author{F. Y. Wang\altaffilmark{1,2}, Y. Y. Wang\altaffilmark{1}, Yuan-Pei Yang\altaffilmark{3}, Y. W. Yu\altaffilmark{4}, Z. Y. Zuo\altaffilmark{5} and Z. G. Dai\altaffilmark{1,2}}
\affil{
$^1$ School of Astronomy and Space Science, Nanjing University, Nanjing 210093, China; fayinwang@nju.edu.cn\\
$^2$ Key Laboratory of Modern Astronomy and Astrophysics (Nanjing University), Ministry of Education, Nanjing 210093, China\\
$^3$ South-Western Institute for Astronomy Research, Yunnan University, Kunming, China; ypyang@ynu.edu.cn\\
$^4$ Institute of Astrophysics, Central China Normal University, Wuhan 430079, China\\
$^5$ School of Science, Xi'an Jiaotong University, Xi'an, 710049, China
}

\begin{abstract}
Young neutron stars (NSs) born in core-collapse explosions are
promising candidates for the central engines of fast radio bursts
(FRBs), since the first localized repeating burst FRB 121102 happens
in a star forming dwarf galaxy, which is similar to the host
galaxies of superluminous supernovae (SLSNe) and long gamma-ray
bursts (LGRBs). However, FRB 180924 and FRB 190523 are localized to
massive galaxies with low rates of star formation, compared with the
host of FRB 121102. Meanwhile, the offsets between the bursts and
host centers are about 4 kpc and 29 kpc for FRB 180924 and FRB
190523, respectively. These properties of hosts are similar to short
gamma-ray bursts \textbf{(SGRBs)}, which are produced by mergers of
binary neutron star (BNS) or neutron star-black hole (NS-BH).
Therefore, the NSs powering FRBs may be formed in BNS mergers. In
this paper, we study the BNS merger rates, merger times, and predict
their most likely merger locations for different types of host
galaxies using population synthesis method. We find that the BNS
merger channel is consistent with the recently reported offsets of
FRB 180924 and FRB 190523. The offset distribution of short GRBs is
well reproduced by population synthesis using galaxy model which is
similar to GRB hosts. The event rate of FRBs (including
non-repeating and repeating), is larger than those of BNS merger and
short GRBs, which requires a large fraction of observed FRBs
emitting several bursts. Using curvature radiation by bunches in NS
magnetospheres, we also predict the observational properties of FRBs
from BNS mergers, including the dispersion measure, and rotation
measure. At late times ($t\geq1$yr), the contribution to dispersion
measure and rotation measure from BNS merger ejecta could be
neglected.
\end{abstract}

%\keywords{radio bursts - magnetars - star: evolution - gamma-ray bursts}

\section{Introduction}
Fast radio bursts (FRBs) are transients of coherent emission lasting
about a few milliseconds \citep{Lorimer07,Thornton13,
Champion16,Shannon18,Platts18,Petroff19,Cordes19} with large
dispersion measures (DMs). Thanks to multi-wavelength follow-up
observations and a precise localization, the repeating FRB 121102
was localized in a dwarf star-forming galaxy at $z= 0.19$
\citep{Chatterjee17,Tendulkar17,Marcote17}. This FRB is also
spatially associated with a persistent radio source
\citep{Chatterjee17,Marcote17}. The properties of this galaxy is
similar to the host galaxies of superluminous supernovae (SLSNe) and
long gamma-ray bursts (LGRBs)
\citep{Tendulkar17,Metzger17,Nicholl17,Zhang19}. This has led to the
hypothesis that bursts are produced by young active neutron stars
(NSs)
\citep{Popov13,Lyubarsky14,Kulkarni14,Katz16,Lu16,mur16,Wang17,Metzger17,Beloborodov17,Lu18,Yang18,wjs19,Beloborodov19,Cheng2020}.
The young active NSs are born in core-collapse explosions. Recently,
a second repeating FRB 180916.J0158+65 was localized to a
star-forming region in a nearby massive spiral galaxy
\citep{Marcote20}. Two coherent mechanisms are often considered in
current FRB models: curvature radiation by bunches
\citep{kat14,kat18,kum17,ghi17,Yang18} and the maser mechanisms
\citep{Lyubarsky14,Beloborodov17,ghi17b,wax17,Lu18,Metzger19,Beloborodov19}.

Recently, two FRBs, which are both single bursts, have been
localized to massive galaxies by the Australian Square Kilometre
Array Pathfinder (ASKAP) and Deep Synoptic Array ten-antenna
prototype (DSA-10) respectively. FRB 180924 \textbf{occured} in a
massive galaxy at redshift $z = 0.32$ with lower star formation rate
\citep{Bannister19}. The offset between the burst and host galaxy
center is about 4 kpc \citep{Bannister19}. The DM contributed by
host galaxy of FRB 180924 is small (between 30-81 pc cm$^{-3}$),
which makes FRBs become a promising cosmological probe
\citep{Gao14,Zhou14,Wei15,Yang16,
Yu17,Walters18,Wang18,Li18,Yu18,zhang18frb,lzx19,liu19}. The Deep
Synoptic Array ten-antenna prototype (DSA-10) localized FRB 190523
to a massive galaxy at a redshift of 0.66 \citep{Ravi19}. This
galaxy is different from the host of FRB 121102, as it is about a
thousand times more massive, with a specific star formation rate two
orders of magnitude lower \citep{Ravi19}. Aside from the two
confirmed host galaxies, \citet{li2019} recently found that the host
candidates of some nearby FRBs are also not similar to that of FRB
121102. This low star formation rate and large offset of FRB hosts
indicate that the NSs powering FRBs may be formed by mergers of
binary neutron stars (BNSs), which is similar to short GRBs.
Observationally, short GRBs occur in early-type galaxies with low
star formation \citep{Barthelmy05,Gehrels05,Bloom06} and have large
offsets from the centers of the host galaxies \citep{Fox05,Fong13}.
Some short GRBs show X-ray flares and internal plateaus with rapid
decay at the end of the plateaus, which are consistent with a
millisecond NS born in BNS merger
\citep{Dai06,Metzger08,Rowlinson13,Wang13,Lu15}.

There are at least two distinct NSs formation channels, i.e.,
core-collapse explosions and BNS mergers. The BNS merger channel is
different from the core-collapse explosion channel in two aspects.
First, the ejecta produced by BNS merger has higher velocity ($v\sim
(0.1-0.3) c$) and lower mass (($10^{-4}-10^{-2}) M_\odot$), compared
with that of core-collapse explosion. The ejecta can affect the
observational properties of FRBs \citep{Margalit19}. Second, the
offsets in the two cases are different. At the time of birth, NSs
receive natal kicks, which are connected to asymmetries in supernova
explosions. Due to the long delay time, BNS will merge at large
radius in host galaxies even if born at small radius
\citep{Bloom99,Fryer99,Belczynski06}. However, in our population
synthesis, most of primary neutron stars receive small kicks (see
Sec. \ref{kick velocity} for details). \cite{Margalit19} also
studied the properties of FRBs from magnetars born in BNS mergers
and accretion induced collapse.

In this paper, we perform an updated analysis of BNS mergers using
population synthesis methods
\citep{Bloom99,Fryer99,Belczynski02,Perna02,Voss03,Belczynski06} and
calculate the FRB properties from BNS mergers. Compared with
previous population synthesis codes \citep{Hurley02,Belczynski08},
the new version of binary star evolution ($\mathtt{BSE}$) code
\citep{Banerjee19} includes several major upgrades. The most
important one is the new formation channel of NSs from
electron-capture supernova (ECS). For close binaries, stars with
masses between $(6-8)~M_\odot$ can form ECS-NSs
\citep{Podsiadlowski04}. We also show the properties of FRBs
powering by NSs from BNS mergers using curvature radiation by
bunches \citep{Yang18}.

The paper is structured as follows. The description of
binary-star-evolution (BSE) code, galaxy potential models, and kick
velocity are presented in section \ref{BSE}. The results of
population synthesis, including merger time and delay time
distributions, offsets distribution and merger rate, are given in
section \ref{sec:Result}. In section \ref{sec:rate}, we compare the
event rates of FRBs, short GRBs and BNS mergers. The observational
properties of FRBs from NS formed by BNS mergers are shown in
section \ref{sec:FRB}. Conclusions and discussion are given in
section \ref{sec:conclusions}.

\section{Compact binaries from population synthesis}\label{BSE}
\subsection{The $\mathtt{BSE}$ code}
The $\mathtt{BSE}$ code, developed by Jarrod Hurley, Onno Pols and
Christopher Tout, is a rapid binary-evolution algorithm based on a
suite of analytical formulae \citep{Hurley02}. It is incorporated
into the N-body evolution program $\mathtt{NBODY7}$
\citep{Aarseth12} for globular cluster as the stellar-evolutionary
sector. Three major upgrades are added to the new version of
$\mathtt{BSE}$ code \citep{Banerjee19}, including (i)  the
semi-empirical stellar wind prescriptions \citep{Belczynski10}, (ii)
remnant formation and material fallback \citep{Fryer12} and the
occurrences of pair-instability supernova (PSN) and pulsation
pair-instability supernova (PPSN) \citep{Belczynski16}, and (iii) a
modulation of the BHs' and the NSs' natal kicks based on the
fallback fraction during their formation \citep{Banerjee19}.

The new $\mathtt{BSE}$ code includes the electron-capture-supernovae
(ECS)-NS formation channel \citep{Podsiadlowski04,Belczynski08}
compared to the previous version \citep{Hurley02}. The primary star
in binary system with initial mass in the range $6-8~M_\odot$ is
likely to become an electron-capture supernova
\citep{Podsiadlowski04}, producing the ECS-NS, which typically has
so small ($<$ 20 km/s) or zero kick velocity that remains bound to
globular clusters whose escape velocities are 10--20 km/s
\citep{Katz75}. These NSs are distinctly least massive NSs,  born
with characteristic mass $m_{\mathrm{ECS}, \mathrm{NS}}=1.26
M_{\odot}$ \citep{Banerjee19}.

Besides, we adopt the remnant-mass prescription of
\cite{Belczynski02} ($\mathtt{nsflag} = 1$), no PPSN/PSN mass cutoff
($\mathtt{psflag} = 0$) and standard, momentum-conserving kick of
\cite{Belczynski08} ($\mathtt{kmech} = 1$). The program
$\mathtt{popbin}$ is used to carry out the population synthesis.

\subsection{Parameter distribution}
We created a catalog of 1000,000 binary systems in which the the
initial system parameters ($M_1, q, e, P$) satisfy the following
distributions
\begin{equation}
f_{M_{1}}\left(M_{1}\right) \propto {M_{1}^{-\alpha},}\qquad {\text
{ for } M_{1} \in[5,100]}
\end{equation}
\begin{equation}
f_{q}(q) \propto q^{\kappa}, \qquad \text { for } q \in[0.1,1]
\end{equation}
\begin{equation}
f_{e}(e) \propto e^{\eta}, \qquad \text { for } e \in[0.0,1.0]
\end{equation}
\begin{equation}
f_{P}(\log_{10} P) \propto(\log_{10} P)^{\pi}, \qquad \text { for }
\log_{10} P \in[0.15,5.5]
\end{equation}
where $M_1$ is the mass of the primary star, $\alpha = 2.7$
\citep{Scalo86}; $q\equiv m_2/m_1$ is the mass ratio of the two
stars; $P$ and $e$ are orbital period and eccentivity respectively;
$\kappa = 0$ \citep{Bethe98}, $\eta= 1$ \citep{Duquennoy91} and $\pi
= 0.5$ \citep{Sana12} are used in our simulation. It is worth
mentioning that the initial binary properties do not significantly
affect (within a factor of 2) the predictions of double compact
object merger rates \citep{deMink15}. The metallicity $Z$ and
maximum evolution time $T$ are set to 0.02 and $15,000\megayear$ for
all binaries.

\subsection{Motion of binaries in the gravitational field of the galaxy}
To obtain the predicted offset of the binary compact objects merger,
we need to know the motion of the binary, which depends on its
initial locations and velocity, the gravitational field of the
galaxy, the kick velocity as well as the delay time. Here, we
consider the spiral galaxy and elliptical galaxy with different
sizes.

For a spiral galaxy, there are three components: a disk, a bulge and
a halo. The galactic disk and bulge potential was proposed by
\cite{Miyamoto75}
\begin{equation}
    \Phi_i(R, z) = - \frac{G M_{i}}{\sqrt{R^{2} + \left(a_{i} + \sqrt{z^{2} + b_{i}^{2}}\right)^{2}}},
\end{equation}
where $\Phi_1$ and $\Phi_2$ refer to bulge and disk potential
respectively. The bulge potential $\Phi_1$ is described by $M_1 =
1.0\times 10^{10} M_\odot$, $a_1 = 0$ and $b_1 = 0.267\kiloparsec$;
the disk potential $\Phi_2$ is described by $M_2 = 6.5\times 10^{10}
M_\odot$, $a_2 = 4.4\kiloparsec$ and $b_2 = 0.308\kiloparsec$
\citep{Bajkova17}.

For an elliptical galaxy, only bulge and halo are considered. The
bulge potential model is Hernquist model \citep{Hernquist90}
\begin{equation}
    \Phi(r) = -\frac{GM_e}{r+a_e},
\end{equation}
where $M_e = 5\times 10^{11} M_\odot$ and $a_e = 5\kiloparsec$
\citep{Belczynski06}.

The halo potential for both spiral galaxy and elliptical galaxy is
given by Navarro-Frenk-White (NFW) profile \citep{Navarro96}
\begin{equation}
    \Phi_\mathrm{h}(r) = -\frac{GM_\mathrm{h}}{r}\ln\bigg(1+\frac{r}{a_\mathrm{h}}\bigg),
\end{equation}
where $M_h = 2.9\times 10^{11} M_\odot$ and $a_h = 7.7\kiloparsec$
are provided by \cite{Bajkova17}.

\subsection{Initial conditions}
The cylindrical coordinates $(R_0, z_0)$ of initial location of the
binary obey the following distributions
\begin{equation}
    p_R(R) \propto R \mathrm{e}^{-R / R_{\mathrm{exp}}},
\end{equation}
\begin{equation}
    p_z(z) \propto \mathrm{e}^{-|z| / z_{\mathrm{exp}}},
\end{equation}
where $R_{\mathrm{exp}} = 4.5 \kiloparsec$, $R_{\mathrm{max}} = 20
\kiloparsec$, $z_{\mathrm{exp}} = 75 \parsec$ and $z_{\mathrm{max}}
= 300 \parsec$ is used \citep{Paczynski90}. For a galaxy with
different size, a scale factor $\alpha$ is used to change the mass
and spatial size proportionally, i.e., $M' = \alpha M$, $R' =
\alpha^{1/3} R$ \citep{Belczynski02}, where $M$ and $R$ are typical
mass and spatial size of a Milky-Way-like galaxy. In this study,
$\alpha = 1,0.1,0.01,0.001$ are considered.

The initial velocity $v_0$ is the local rotational velocity of the
galaxy which has no vertical component \citep{Belczynski02}
\begin{equation}
    v_0 = v_{\phi}(R) = \sqrt{\frac{GM(<R)}{R}}
\end{equation}
where $M(<R) = \int_{-\infty}^{+\infty}\int_{0}^{R}2\pi R\rho(R,
z)\ud R\ud z$ is the total mass within a cylinder of radius $R$,
$\rho(R, z)$ is the density of the galaxy as a function of $R$ and
$z$.

\subsection{Equations of motion}
The binary's motion trajectory in the galactic potential field $\phi$ can
be obtained by solving the following equations
\begin{eqnarray}
\frac{\ud x}{\ud t}=v_{x}, \qquad
\frac{\ud y}{\ud t}=v_{y}, \qquad \frac{\ud z}{\ud t}=v_{z}, \qquad \\ \nonumber
\frac{\ud v_{x}}{\ud t} = -\left(\frac{\partial \phi}{\partial x}\right)_{y, z}, \qquad
\frac{\ud v_{y}}{\ud t} = -\left(\frac{\partial \phi}{\partial y}\right)_{x, z}, \qquad
\frac{\ud v_{z}}{\ud t} = -\left(\frac{\partial \phi}{\partial z}\right)_{x, y}
\end{eqnarray}
where $x_0$, $y_0$, $v_{x0}$, $v_{y0}$ are obtained by projecting
$R_0$ and $v_0$ into $x$ and $y$ directions
\begin{eqnarray}
    x_0 = R_0 \cos(l_{\mathrm{bin}}),\qquad y_0 = R_0 \sin(l_{\mathrm{bin}})\\
    v_{x0} = v_0 \cos(l_{\mathrm{bin}} + \frac{\pi}{2}),\qquad v_{y0} = v_0 \sin(l_{\mathrm{bin}} + \frac{\pi}{2})
\end{eqnarray}
where $l_{\mathrm{bin}}$ is randomly sampled galactic longitude of
the binary system. These six motion equations are different from
those in cylindrical coordinates proposed by \cite{Paczynski90}, in
which the signs of $\Phi_i(R, z)$, $\bigg(\dfrac{\partial
\Phi}{\partial R}\bigg)_z$ and $\bigg(\dfrac{\partial \Phi}{\partial
z}\bigg)_R$ are wrong. Solving equations in Cartesian coordinate
system can avoid the difficulty of treating $R$ and $v_R$'s signs
when binaries move across the galactic center.

\subsection{Kick velocities}
\label{kick velocity} NSs formed without any fallback receive full
natal kicks $V_{\mathrm{kick}}$, which follow a Maxwellian
distribution with dispersion $\sigma =  190\kilometer\second^{-1}$
\citep{Hansen97,Hurley02}
\begin{equation}
    f(v)\propto  \frac{v^{2} e^{-v^{2} /\left(2 \sigma^{2}\right)}}{\sigma^{3}}.
\end{equation}

\textbf{In the code model of partial fallback case}, the kick velocity $v_{\mathrm{kick}}$
is modified by the factor $(1 - f_{\mathrm{fb}})$
\begin{equation}
    v_{\mathrm{kick}} = V_{\mathrm{kick}}\left(1-f_{\mathrm{fb}}\right),
\end{equation}
where $f_{\mathrm{fb}}$ is the fraction of the stellar envelope that
falls back. It is worth mentioning that this does not apply to all
NSs. For ECS-NSs, the natal kicks follow a Maxwellian distribution
with a small dispersion and are exempted from the fallback treatment
\citep{Banerjee19}. We assume the kick velocity of the second NS
follows a Maxwellian distribution with $\sigma =
190\kilometer\second^{-1}$ and the direction is perpendicular to the
binary's velocity before the kick. Velocity additions are conducted
to get the new velocities after the first kick and the second kick
respectively. Then the motion of the binary can be divided into
three parts by the time of first kick and the second kick. Each new
velocity is used as the initial velocity for the next stage of
motion to calculate the final offset.

\section{Results of population synthesis}
\label{sec:Result}
%\subsection{Compact binaries merger from population synthesis}
From the population synthesis, we obtain 5531 NS-NS mergers out of
10,000,000 binaries. For NS-NS systems, about $60\sim 70$\% of the
first NSs are ECS-NSs, which have zero or low kick velocities
($\sim$ few $\kilometer\second^{-1}$). It is worth mentioning that
we consider the NS-NSs that can give birth to stable NSs, which are
possible FRB progenitors
\citep{Popov13,Lyubarsky14,Kulkarni14,Katz16,Metzger17,Beloborodov17,Lu18,Yang18}.
The maximum mass of NS depends on the equation of state and spin
period, which are still uncertain. We choose an rough upper limit as
$m_1 + m_2 \leqslant 2.6 M_{\odot}$ or $3 M_{\odot}$. We find 297 and 4523 possible NSs for the
$m_1 + m_2 \leqslant 2.6 M_{\odot}$ and $3 M_{\odot}$ before merger respectively.

\subsection{Merger time and delay time distributions}
\begin{figure}
    \centering
    \includegraphics[width=0.6\textwidth]{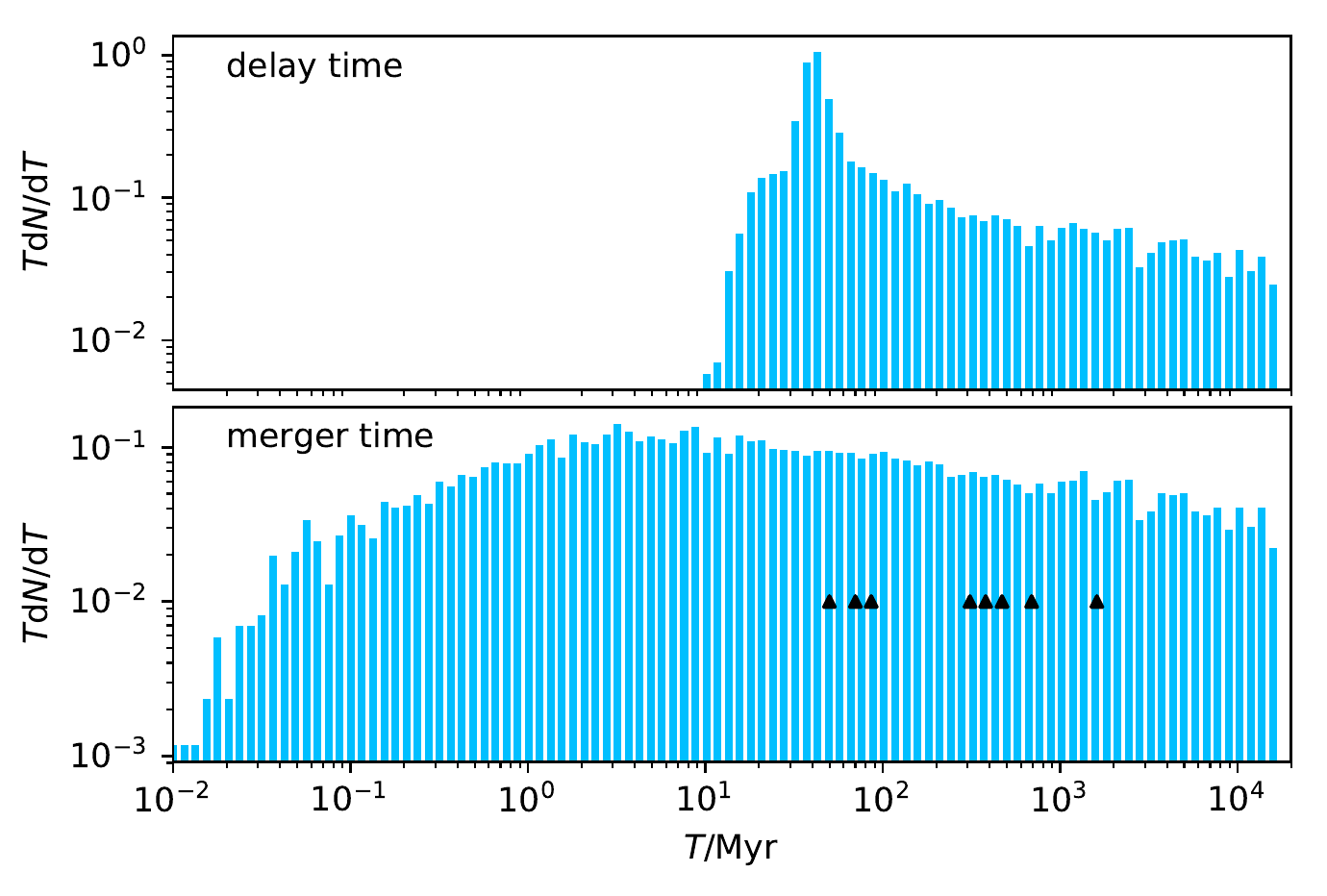}
    \caption{Merger time and delay time distributions for NS-NS systems. The eight Galactic NS-NS systems are shown with triangles.}
    \label{Merger time and delay time: BNS}
\end{figure}

\begin{figure}
    \centering
    \includegraphics[width=0.6\textwidth]{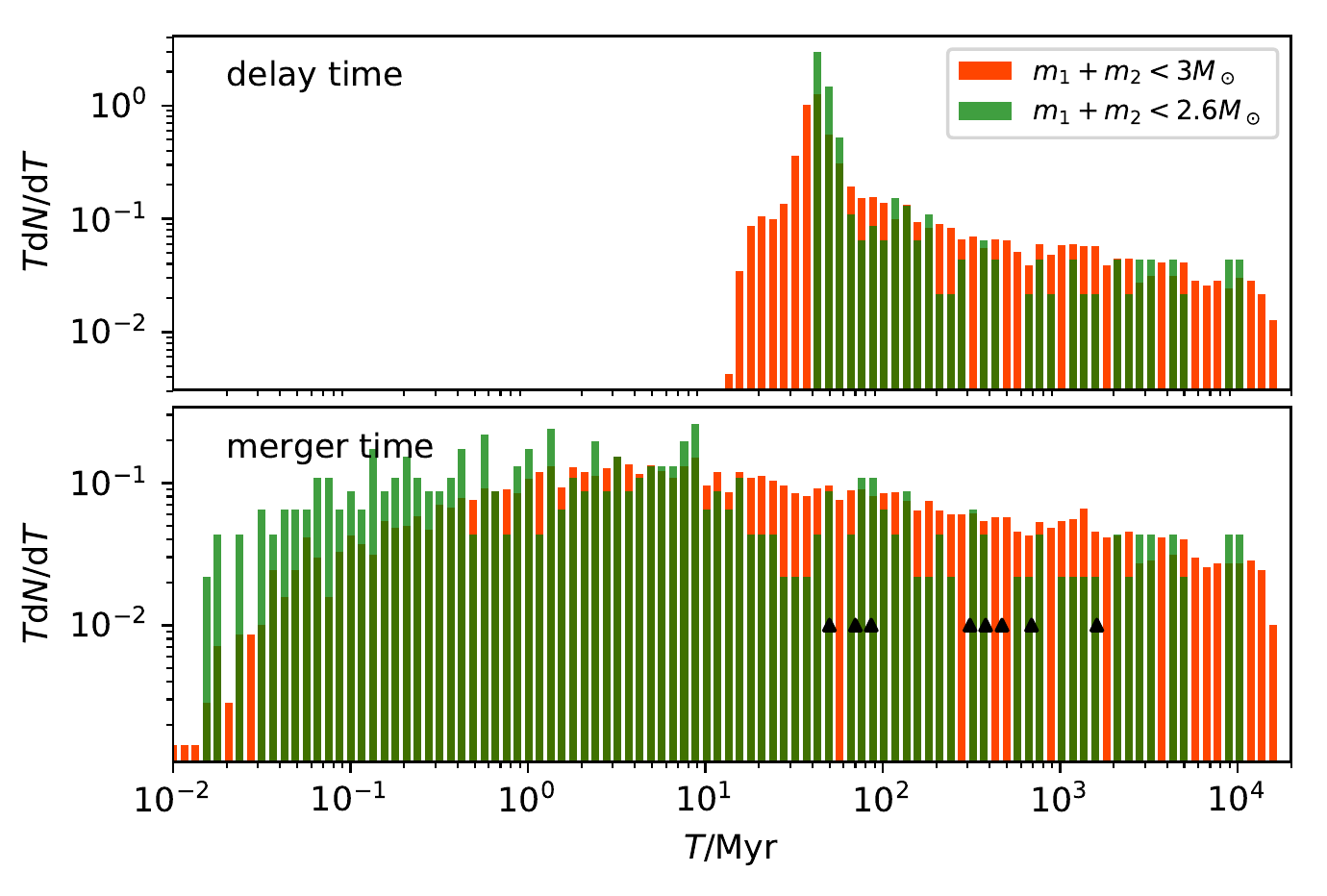}
    \caption{Merger time and delay time distribution for NS-NS systems whose mergers may produce stable NSs.}
    \label{Merger time and delay time: BNS magnetar}
\end{figure}

\begin{figure}
    \centering
    \includegraphics[width=0.6\textwidth]{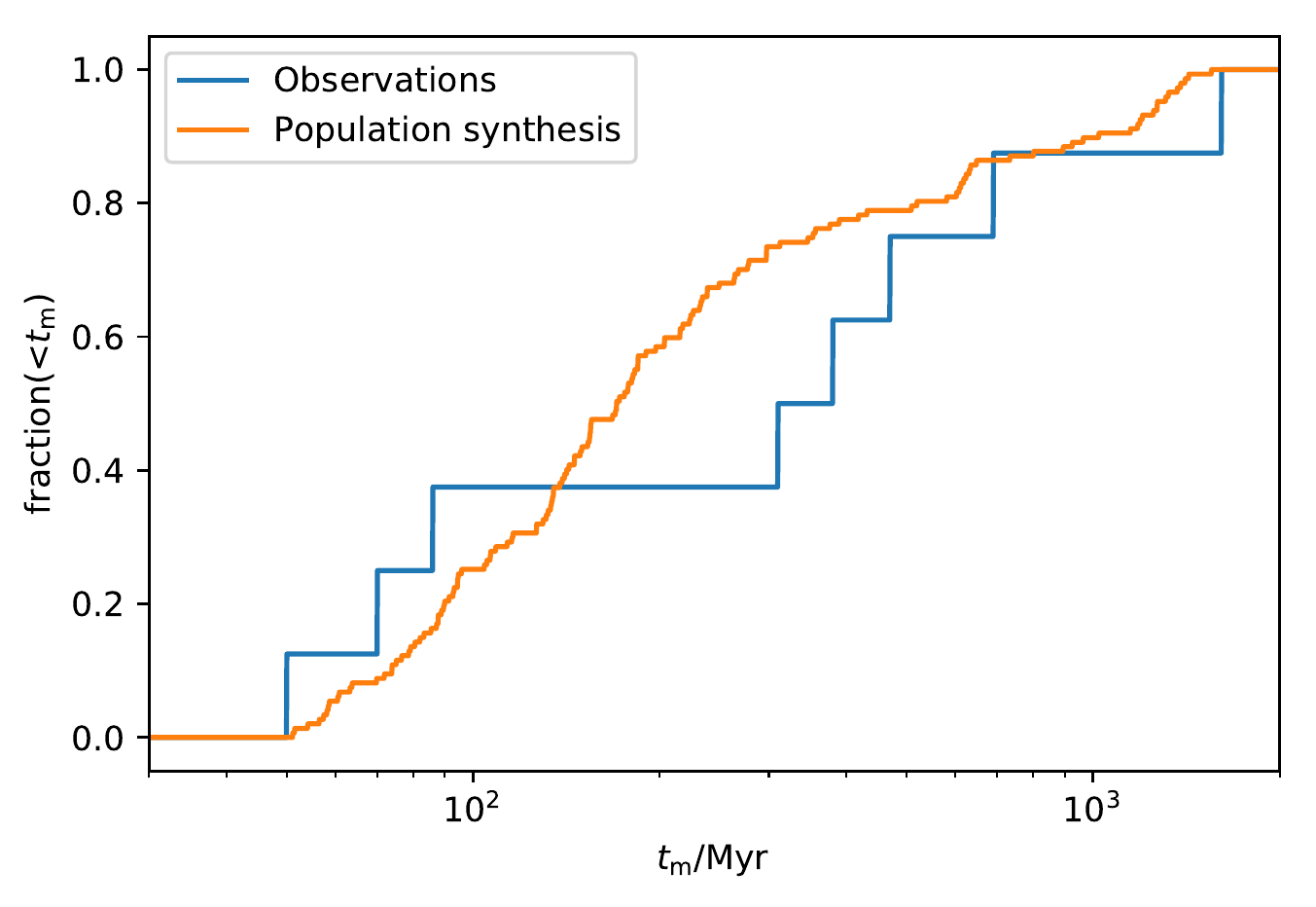}
    \caption{Cumulative distribution of merger times for NS-NS from population synthesis (orange) and observations (blue). The $p$ value is 0.22 from Kolmogorov-Smirnov test.}
    \label{Merger time}
\end{figure}

The delay time is defined as the time between the birth of the
binary system and the final merger, while merger time refers to the
time interval between the binary compact object formation and the
merger. In Figure \ref{Merger time and delay time: BNS}, we show the
delay time distribution and merger time distribution for NS-NS
binaries. The merger times of 8 observed field Galactic NS-NS
systems \citep{Beniamini19} are shown as triangles. We also compare
the cumulative distributions of merger time from population
synthesis with these 8 observed field Galactic NS-NS systems. The
result is shown in Figure \ref{Merger time}. From Kolmogorov-Smirnov
test, the $p$ value is 0.22, which supports that they follow the
same distribution. For the delay time, as NSs' progenitors are
massive stars, their birth rate follows the star formation rate
(SFR) with a minimal delay. The delay time is dominated by this
gravitational wave (GW) insprial time. The time till merger depends
on the initial semimajor axis $a$, and the eccentricity $e$ of the
BNS as $t_\mathrm{m}\propto a^4(1-e)^{7/2}$. Under some assumptions,
the delay time distribution is $\ud N/\ud t\propto t^{-1}$ at late
times ($t\geq1$Gyr) \citep{Piran92,Totani08}. From Figure
\ref{Merger time and delay time: BNS}, it is obvious that the merger
times from population synthesis show a similar distribution. Figure
\ref{Merger time and delay time: BNS magnetar} shows the merger time
and delay time distributions for NS-NS mergers that may produce
stable NSs. We can see that the distributions of delay time and
merger time are almost the same in Figures \ref{Merger time and
delay time: BNS} and \ref{Merger time and delay time: BNS magnetar}.

\subsection{Offset cumulative distribution}
\begin{figure}
    \centering
    \includegraphics[width=1.0\textwidth]{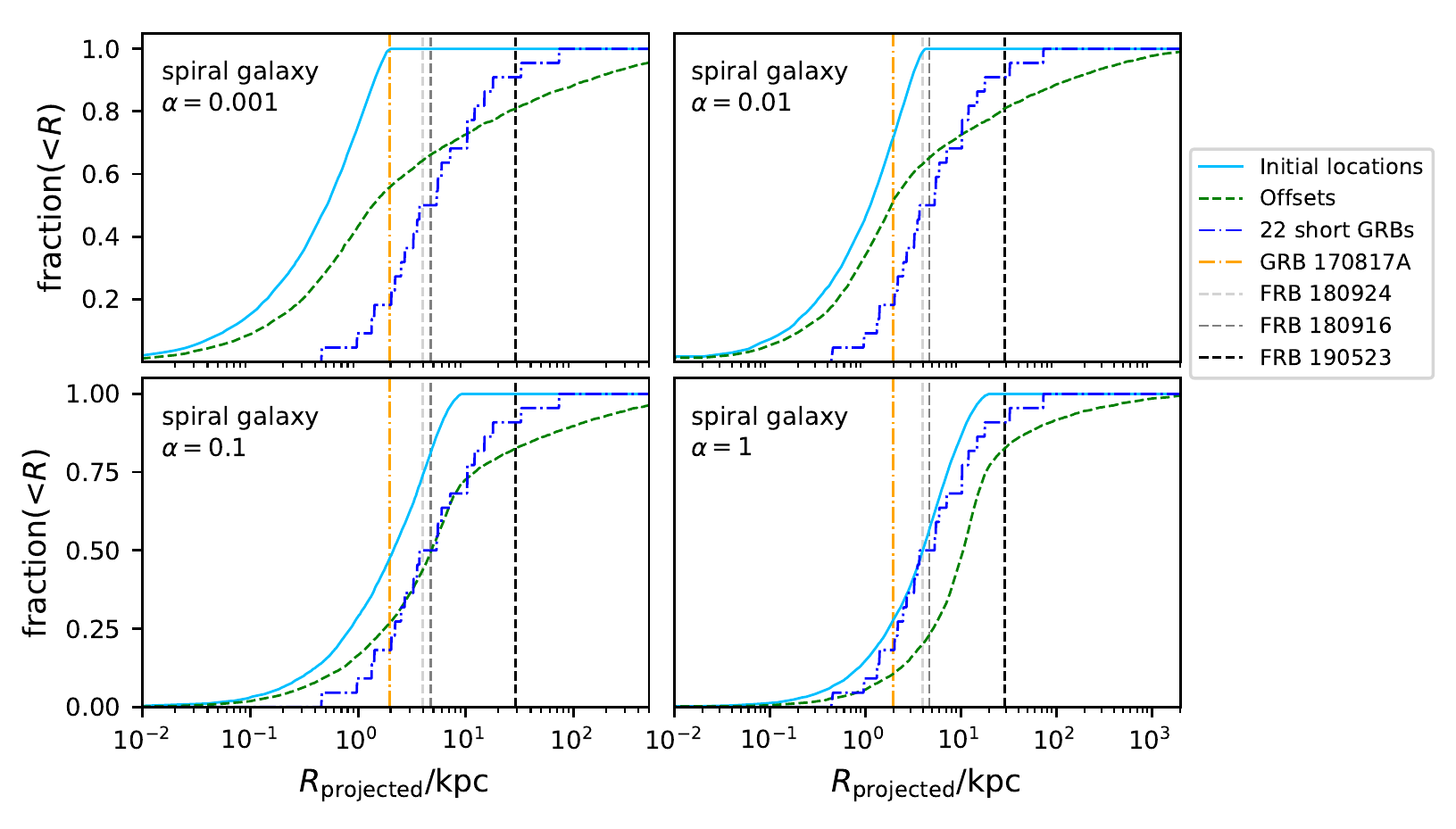}
    \caption{Initial location and offset cumulative distributions in spiral galaxies with different masses ($\alpha=0.001,0.01,0.1,1$) for NS-NS mergers. The observed offsets of 22 short GRBs, GW170817/GRB 170817A, FRB 121102, FRB 180916, FRB 180924 and FRB 190523 are also shown.}
    \label{Offsets:BNS spiral}
\end{figure}

\begin{figure}
    \centering
    \includegraphics[width=1.0\textwidth]{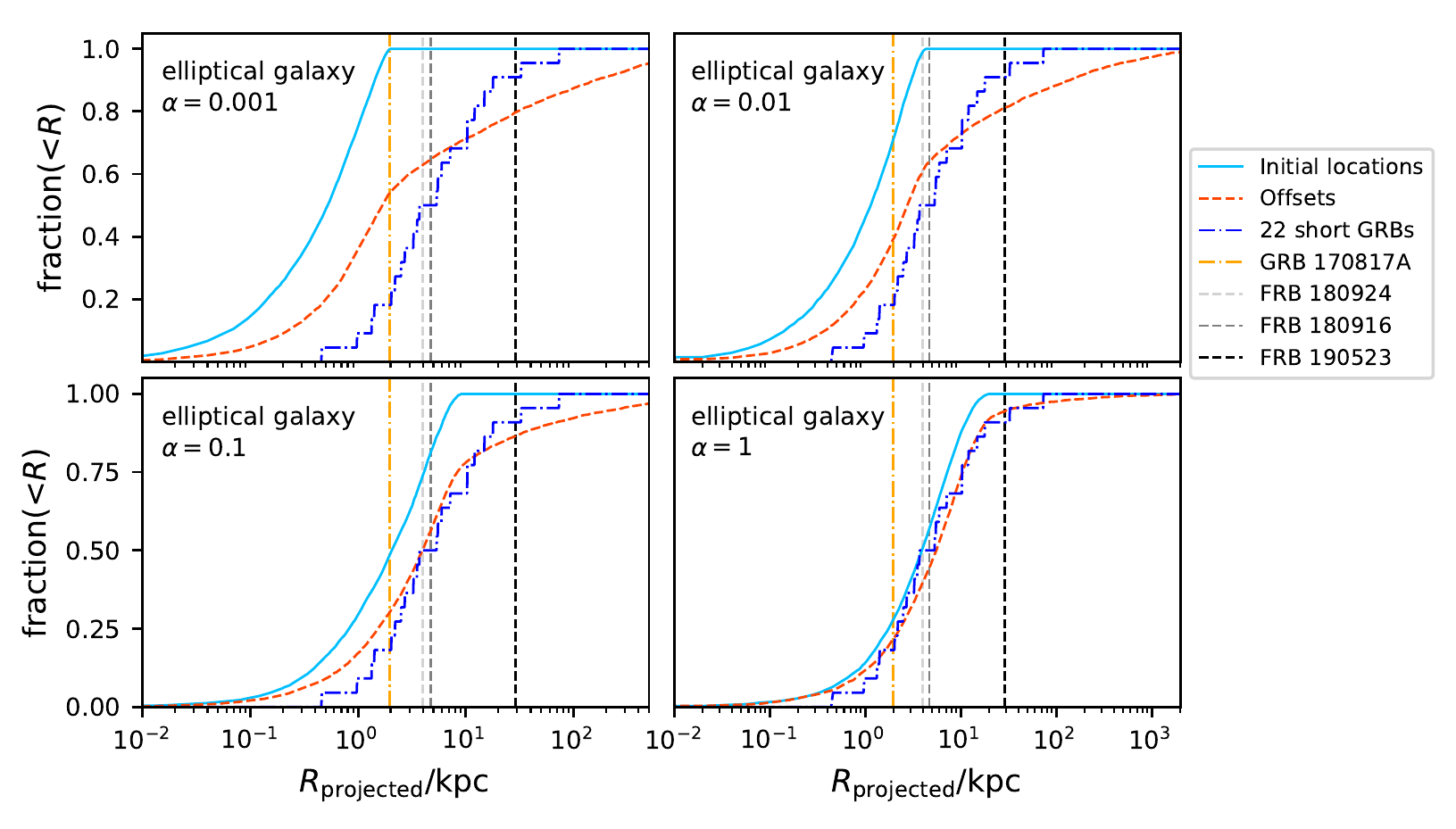}
    \caption{Initial location and offset cumulative distributions in elliptical galaxies with different masses ($\alpha=0.001,0.01,0.1,1$) for NS-NS mergers. The observed offsets of 22 short GRBs, GW170817/GRB 170817A, FRB 121102, FRB 180916, FRB 180924 and FRB 190523 are also shown.}
    \label{Offsets:BNS elliptical}
\end{figure}

The cumulative distributions of offsets between merger locations and
centers of host galaxies for NS-NS mergers are shown in Figure
\ref{Offsets:BNS spiral} and Figure \ref{Offsets:BNS elliptical} for
spiral galaxies and elliptical galaxies, respectively. The projected
offset is the offset in the direction perpendicular to the line of
sight. In the calculations, we average over all possible
orientations of host galaxies. The observed offsets of GW170817/GRB
170817A (2 kpc) \citep{Levan17}, FRB 180924 (4 kpc)
\citep{Bannister19}, FRB 180916 (4.7 kpc) \citep{Marcote20} and FRB 190523 (29 kpc) \citep{Ravi19} and
offsets distribution of short GRBs
\citep{Fong10,Fong13,Berger14} are also shown for comparison. For
massive spiral galaxies ($\alpha=1$), 60\% NS-NS systems will have
offsets larger than 10 kpc. While for low-mass spiral galaxies
($\alpha=0.001,0.01,0.1$), the fraction is about 30\%. For
elliptical galaxies, the cumulative distributions of offsets are
steeper when galaxy mass increases. For massive elliptical galaxies
($\alpha=0.1, 1$) in Figure \ref{Offsets:BNS elliptical}, the
observed offsets of short GRBs are consistent with simulated
NS-NS systems. The observed median mass of short GRBs host
galaxies is about $10^{10}M_\odot$ \citep{Berger14}, which
corresponds to the $\alpha=0.1$ case.

\begin{figure}
    \centering
    \includegraphics[width=0.6\textwidth]{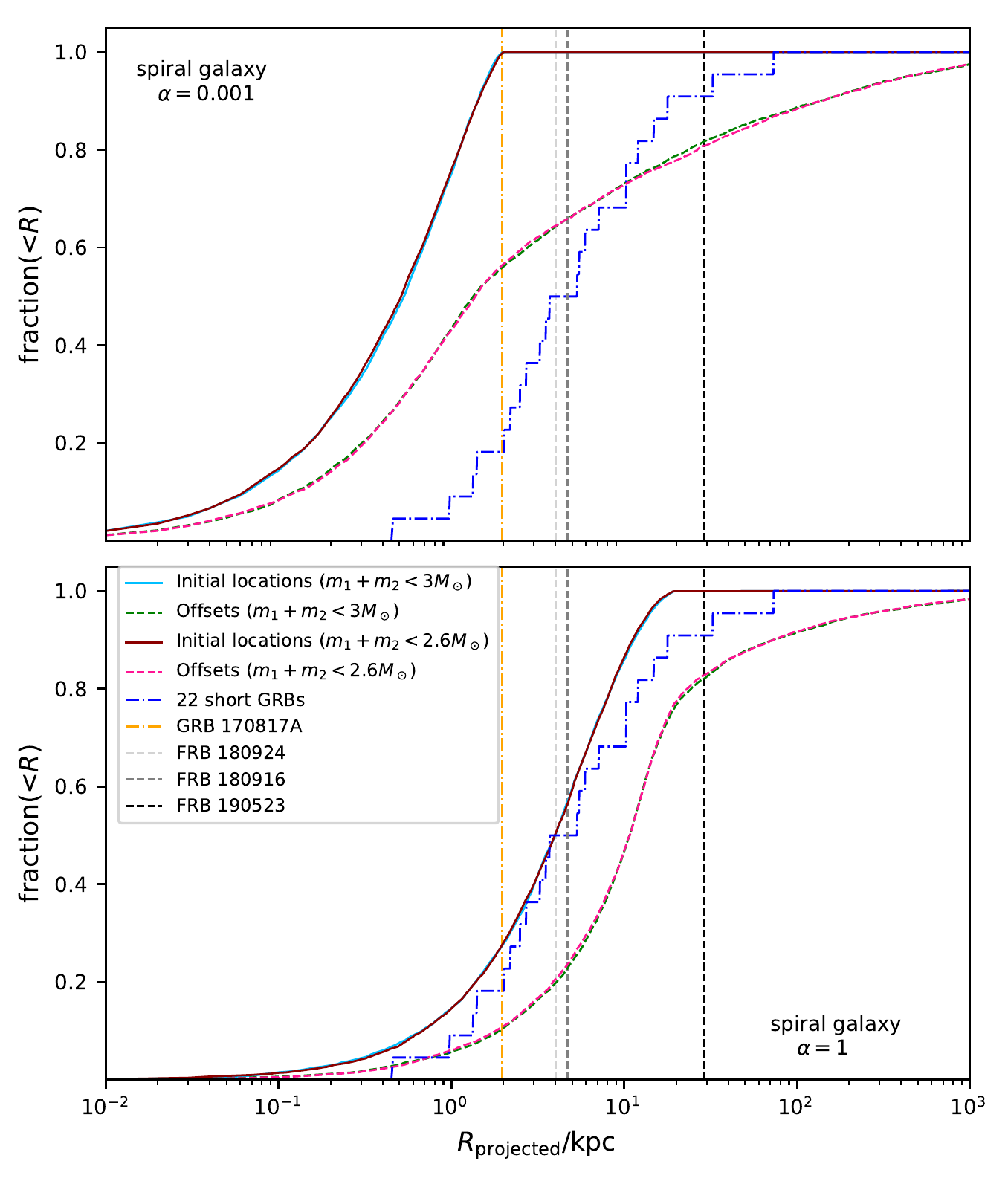}
    \caption{Initial location and offset cumulative distributions in spiral galaxies for NS-NS mergers that may produce NSs and observed offsets of short GRBs and FRBs. }
    \label{Offsets:BNS magnetar spiral}
\end{figure}

Figures \ref{Offsets:BNS magnetar spiral} and \ref{Offsets:BNS
magnetar elliptical} show the offset distributions for NS-NS mergers
that may produce NSs in spiral and elliptical galaxies
respectively. For massive galaxies, the offset distributions for
different upper limits of newborn NS are almost the same. For
low-mass galaxies ($10^{8}M_\odot$), about 70\% NS-NS systems will
merge with offsets less than 5 kpc for $m_1 + m_2 \leqslant 3.0
M_{\odot}$ case, while more than 80\% for $m_1 + m_2 \leqslant 2.6
M_{\odot}$ case.

\subsection{Merger rate}
The merger rate $R_{\mathrm{m}}(z)$ is a convolution of the star
formation rate history $\rho(z)$ and the probability density
function (PDF) of delay time
\begin{equation}
    R_{\mathrm{m}}(z)=\int_{t(z)}^{t(z=\infty)} f \rho(z)\left(t^{\prime}\right) p\left[t(z)-t^{\prime}\right] \ud t^{\prime},
\end{equation}
where $\ud t = - H(z)^{-1}(1+z)^{-1}  \ud z$, $H(z)$ is the Hubble parameter as a function of $z$, $f$ is the mass
fraction of the compact binaries (NS-NS) to the entire stellar
population, $t(z)$ is the cosmic age at redshift $z$. The cosmic
star formation rate (CSFR) is taken from \cite{Madau14}
\begin{equation}
    \rho(z) = 0.015 \frac{(1+z)^{2.7}}{1+[(1+z) / 2.9]^{5.6}} M_{\odot} \text { year }^{-1} \mathrm{Mpc}^{-3}.
\end{equation}
When calculating the integration, we ignore the factor $f$, just
showing the shape of merger rate as a function of redshift. The
merger rate $R_{\mathrm{m}}(z)$ as a function of redshift $z$ are
plotted in Figure \ref{Merger rate: BNS} for $m_1 + m_2 \leqslant 3
M_{\odot}$ before merger.

\section{Local event rate}
\label{sec:rate}

Mergers of NS-NS and NS-BH binaries have long been considered as the
progenitors of short GRBs, which was confirmed by the discovery of
GW170817/GRB 170817A \citep{Abbott17}. According to GW170817, the
local rate of NS-NS mergers is $\rho_0=1540^{+3200}_{-1220} \rm
Gpc^{-3} yr^{-1}$ \citep{Abbott17}. The local rate of short GRBs has
been widely studied and found to range from several to several tens
of $\rm Gpc^{-3} yr^{-1}$ \citep{Guetta06,
Nakar06,Coward12,Wanderman15,Tan18,Zhang18}. According to
\cite{Zhang18}, the local rate of short GRBs is 7.53 $\rm Gpc^{-3}
yr^{-1}$. If the beaming factor is chosen as $27^{+158}_{-18}$
\citep{Fong15}, the local event rate of all short GRBs is
$203_{-135}^{+1152} \rm Gpc^{-3} yr^{-1}$, which is broadly in
agreement with the LIGO result.

As for the formation rate of non-repeating FRBs, we adopt the model
and results given by \cite{Zhang19}. In their calculation,
considering the delay time, the cumulative redshift distribution can
be derived as
\begin{equation}
N(<z) = T\frac{A}{4\pi}f \int_0^z \rho_{\mathrm{FRB}}(z)[\int_0^1
\eta(\varepsilon)\int_{E_th/\varepsilon}^{E_{max}}\Phi(E)\ud E\ud\varepsilon]
\times \frac{\ud V(z)}{1 + z},
\end{equation}
where $ T $ is the observation time, $ A $ is the sky area, $ f_b $
is the beaming factor of FRB, $ \Phi(E) $ is the energy
distribution, $ \eta(\varepsilon) $ is the beaming effect of
telescope and $ \rho_{FRB}(z) $ is the formation rate of FRBs. We
adopt the results of Parkes with time delay \citep{Zhang19} and
obtain the local formation rate as
\begin{equation}
\rho_0 \simeq 6\times 10^4\, \bigg(\frac{T}{270\mathrm{s}}\bigg)^{-1}
\bigg(\frac{A}{4\pi}\bigg)^{-1}f_b^{-1} \mathrm{Gpc}^{-3} \mathrm{yr}^{-1},
\end{equation}
where the typical value of observation time is take from
\citet{Thornton13}. This local formation rate is consistent with the
result of \cite{Cao18}. It is proposed that, during the final stages
of a BNS merger inspiral, the interaction between the magnetosphere
could also produce a non-repeating FRB
\citep{Totani13,Zhang14,Wang16,Metzger16,wjs18,Zhang20}. However,
the rate of BNS merger is well below the FRB rate, which is also
discussed in \cite{Ravi19b} for CHIME sample.

On the other hand, many bursts can be produced by the remnant NSs
from BNS mergers \citep{Yamasaki18}. If the life time of each
repeater is $\tau$ years, the volume density $\rho_{\rm FRB}\tau
\sim 5\times10^3\, \rm{Gpc}^{-3}$ is found by \cite{Wang19} above
the fluence limit $F_{\rm min}=0.5$ Jy ms. The value of $\tau$ is
very uncertain, which relates to the activity of newborn neutron
star. The magnetic activity timescale in the direct Urca (high-mass
NS) and modified Urca (normal-mass NS) case are about a few ten
years and a few hundred years, respectively \citep{Beloborodov16}.
For a lifetime of $\tau\sim (10-100)$ years, the birth rate
$\rho_{\rm FRB} \sim (50-500)~\rm{Gpc}^{-3}yr^{-1}$ can explain the
observational properties of FRB 121102, such as energy distribution
\citep{Wang19}. In this case, the observed FRB rate is consistent
with those of SLSNe and SGRBs \citep{Wang19}.

\section{Observational properties of FRBs from BNS mergers}
\label{sec:FRB}

Most FRB models invoke coherent radiation within the magnetosphere
of a magnetized neutron star \citep[e.g.][]{kum17,Yang18,wjs19}.
%In the pulsar magnetosphere,
Only the fluctuation of net charges with respect to the background
outflow can make a contribution to coherent radiation
\citep{Yang18}. The fluctuation in the magnetosphere could be
triggered by non-stationary sparks \citep{rud75}, pulsar lightning
\citep{kat17}, starquake of neutron star \citep{wwy2018}, a nearby
astrophysical plasma stream \citep{zha17b,zb2018}, or exotic small
bodies \citep{gen15,dai16}. In this section, we calculate the
coherent curvature radiation by the fluctuation of the
magnetosphere, and predict observation properties from the ejecta of
the BNS merger.

\subsection{Coherent curvature radiation by bunches}

The extremely high brightness temperatures of FRBs require
their radiation mechanisms to be coherent. Two coherent mechanisms
are often considered in current FRB models: curvature radiation by
bunches \citep{kat14,kat18,kum17,ghi17,Yang18} and the maser
mechanisms
\citep{Lyubarsky14,Beloborodov17,ghi17b,wax17,Lu18,Metzger19}. Here,
we mainly consider curvature radiation by bunches as the radiation
mechanism of FRBs.

First, we briefly summarize the curvature radiation by bunches
following \citet{Yang18}. For a relativistic electron with Lorentz
factor $\gamma$ moving along with a trajectory with curvature radius
$\rho$, its radiation is beamed in a narrow cone of $\sim1/\gamma$
in the electron velocity direction. In the trajectory
plane, the energy
radiated per unit frequency interval per unit solid angle is
\citep{jac75} \be \frac{\ud I}{\ud\omega
    \ud\Omega}\simeq\frac{e^2}{c}\left[\frac{\Gamma(2/3)}{\pi}\right]^2\left(\frac{3}{4}\right)^{1/3}
\left(\frac{\omega\rho}{c}\right)^{2/3}e^{-\omega/\omega_c},\label{scr2}
\ee where $\omega_c=3c\gamma^3/2\rho$ is the critical frequency of
curvature radiation. We consider a three-dimensional bunch
characterized by its length $L$, curvature radius $\rho$, bunch
opening angle\footnote{Here the bunch opening angle is defined as
the maximum angle between each electron trajectory (i.e., magnetic
field line for curvature radiation) in a bunch, see Figure 9 and
Figure 10 in \cite{Yang18}. In the magnetosphere, considering that
the field lines are not parallel with each other, bunches will
slightly expand when they moves away from the dipole center.}
$\varphi$. In the bunch, the electron energy distribution is assumed
to be \be
N_e(\gamma)\ud\gamma=N_{e,0}(\gamma/\gamma_m)^{-p}\ud\gamma~~~{\rm
    for}~~~\gamma_m<\gamma<\gamma_M. \ee
$N_e$ is defined as the net charge number of electrons in a bunch, considering only
the net charged particles contribute to the coherent curvature
radiation. Then the curvature radiation spectra are characterized by
a multi-segment broken power law \citep[see Figure 11 and Figure 12 in][]{Yang18}, with the break frequencies defined
by \be \nu_l=\frac{c}{\pi
    L},~~~\nu_\varphi=\frac{3c}{2\pi\rho\varphi^3}~~~{\rm
    and}~~~\nu_c=\frac{3c\gamma_m^3}{4\pi\rho}. \label{threefreq} \ee The emitted energy
at the peak frequency $\nu_{\rm peak}$ is given by \citet{Yang18} \be
\left.\frac{\ud I}{\ud\omega \ud\Omega}\right|_{\rm peak} \simeq
\frac{e^2}{c}K(p)N_{e,0}^2\gamma_m^4\left(\frac{\nu_{\rm
        peak}}{\nu_{c}}\right)^{2/3},\nonumber\\\label{fmax} \ee where
$K(p)=2^{(2p-6)/3}\left[\Gamma\left(2/3\right)\Gamma\left((p-1)/3\right)\right]^2/3\pi^2$, and the
peak frequency is given by $\nu_{\rm
    peak}=\min(\nu_l,\nu_\varphi,\nu_{c})$.

For a single source, e.g., a charged particle or bunch, the
frequency-dependent duration of the curvature radiation is
$T\sim(1-v/c)\rho/\gamma c\sim\rho/2c\gamma^3\sim1/\nu_c$. If the
electromagnetic wave frequency is considered to be
$\nu_c\sim10^9~{\rm Hz}$ that corresponds to the typical FRB
frequency, the pulse duration of the curvature radiation will be
$T\sim1/\nu_c\sim1~{\rm ns}$, which is much less than the observed
FRB duration, $T_{\rm obs}\sim 1~{\rm ms}$. Therefore, there must be
numerous bunches sweeping across the line of sight during the
observed duration $T_{\rm obs}$ \citep{Yang18}, as shown in Figure
\ref{fig1}. Assuming that the radiation from numerous bunches is
incoherent at the observed frequency $\nu\sim1~{\rm GHz}$, the
observed flux density satisfies \citep{Yang18} \be
F_{\nu}=\frac{2\pi}{TD^2}\frac{\ud I}{\ud\omega \ud\Omega}, \ee where $D$ is
the source distance, and $T$ is the mean time interval between
adjacent bunches. If the bunches are generated via plasma
instability, the gap between adjacent bunches may be of the order of the
bunch scale itself, which gives $T\sim L/c$. Thus, the observed
peaked flux density is \be F_{\nu,{\rm peak}}\simeq\frac{2\pi
e^2}{c} K(p) \frac{N_{e,0}^2\gamma_m^4}{D^2T}\left(\frac{\nu_{\rm
peak}}{\nu_{c}}\right)^{2/3}. \ee Due to $\nu_{\rm
peak}\leqslant\nu_c$, the observed peak flux density must be less
than the limit flux density \be F_{\nu,{\rm limit}}=\frac{2\pi e^2}{c} K(p)
\frac{N_{e,0}^2\gamma_m^4}{D^2T}.\label{fnul} \ee
%If $\nu_{\rm peak}=\nu_c$, one has $F_{\nu,{\rm peak}}=F_{\nu,{\rm limit}}$

\subsection{FRBs from pulsar magnetosphere}

As pointed out by \citet{Yang18}, only the fluctuation of net
charges with respect to the background outflow can make a
contribution to coherent radiation. We define the fraction between
the fluctuating net charge density $\delta n_{\rm GJ}$ and the
background Goldreich-Julian density $n_{\rm GJ}$ as \be
\mu_c=\frac{\delta n_{\rm GJ}}{n_{\rm GJ}}, \ee one has $N_{e,{\rm
tot}}=\gamma_mN_{e,0}/(p-1)=\mu_cn_{\rm GJ}V$, where $V$ is the
bunch volume. When the bunches move along the field lines in the
pulsar magnetosphere, the bunch opening angle $\varphi$ would
depends on the magnetic field configure, as shown in Figure
\ref{fig1}. If the emission region is close to the magnetic axis, we
consider that $\varphi\sim\theta$, where $\theta$ is the polar
angle. Therefore, the transverse size of a bunch is $\sim
r\varphi\sim \rho\varphi^2$, where $r\sim\rho\theta$ is the bunch
distance related to the center of magnetic dipole\footnote{According
to the geometry of the magnetic dipole field (e.g., see Appendix G
of \citet{Yang18}), the curvature radius at $(r,\theta)$ is
$\rho\simeq4r/3\sin\theta$ for $\theta\lesssim0.5$.}. The bunch
volume is $V\sim L(\rho\varphi^2)^2\simeq L\rho^2\varphi^4$, and one
has \be N_{e,0}=(p-1)\gamma_m^{-1}\mu_cn_{\rm
GJ}L\rho^2\varphi^4\label{ne0} \ee On the other hand, the background
Goldreich-Julian density is \be n_{\rm
GJ}=\frac{B}{Pec}\left(\frac{r}{R}\right)^{-3}\sim\frac{BR^3}{ecP\rho^3\varphi^3},\label{ngj}
\ee where $B$ is the pulsar surface magnetic field strength, $P$ is
the pulsar period, and $R$ is the pulsar radius. According to
Eq.(\ref{fnul}), Eq.(\ref{ne0}), Eq.(\ref{ngj}) and $T\sim L/c$, one
finally has \be
F_{\nu,{\rm limit}}&\simeq&\frac{2\pi(p-1)^2K(p)}{c^2}\frac{\mu_c^2\gamma_m^2 LB^2R^6\varphi^2}{D^2 P^2\rho^2}\nonumber\\
&\simeq&3~{\rm
Jy}\mu_{c}^2\gamma_{m,2}^2L_1B_{14}^2R_6^6\varphi_{-2}^2D_{\rm
Gpc}^{-2}P_{-1}^{-2}\rho_{7}^{-2}, \ee where the convention
$Q_x=Q/10^x$ in units of cgs is adopt, $D_{\rm Gpc}=D/1~{\rm Gpc}$,
and $(p-1)^2K(p)=0.4542$ for $p=3$. The intrinsic spectrum is a
multi-segment broken power law\footnote{In general, for a bunch with
uniform distributed charge density, its curvature radiation appears
a wide spectrum, i.e., $\Delta\nu\sim\nu$ \citep{Yang18}. The
observed structure might be due to scintillation, plasma lensing
\citep{cor17}, or spatial structure of a clumpy radiating charge
distribution \citep{kat18}.} with break frequencies of $\nu_l$,
$\nu_\varphi$ and $\nu_c$ \citep{Yang18}. In the above equation, we
take $L\sim10~{\rm cm}$, $\varphi\sim0.01$, $\gamma_m\sim100$ and
$\rho\sim10^7~{\rm cm}$, which make
$\nu_l\sim\nu_\varphi\sim\nu_c\simeq1~{\rm GHz}$ according to
Eq.(\ref{threefreq}). In this case, the spectrum above a few GHz
would be very soft (with spectral index of $-(2p+4)/3$ as discussed
in \citet{Yang18}), which might explain the high-frequency cutoff of
FRB observation. Meanwhile, if we consider that the observed narrow
spectrum is due to very-soft spectrum at high frequency and radio
absorption at low frequency (free-free absorption, synchrotron
self-absorption, plasma absorption and etc.), the condition of the
above typical frequencies at a few GHz would make the limit flux is
minimum, which would gives a strongest constraint on these
parameters.
%Finally one has $F_{\nu,{\rm peak}}\sim F_{\nu,{\rm limit}}\simeq 3~{\rm Jy}$.
For an FRB with observed flux density of $F_\nu\sim{\rm
    a~few}~{\rm Jy}$, its progenitor could be a young magnetar with
$B\gtrsim10^{14}~{\rm G}$ and $P\lesssim0.1~{\rm s}$, which could
cause a large observed flux density limit $F_{\nu,{\rm limit}}$.

After the BNS merger, the newborn neutron star with a surface
dipole magnetic field strength $B$ and initial period $P_0$ would
spin down due to magnetic torques, the spindown luminosity $L_{\rm
sd}$ is given by \citep[e.g.][]{yan19} \be L_{\rm sd}=L_{\rm
sd,0}\left(1+\frac{t}{t_{\rm
        sd}}\right)^{-2}\simeq10^{43}~\unit{erg~s^{-1}}\left\{
\begin{aligned}
    &0.4B_{14}^{-2}t_{\rm yr}^{-2}&\text{for}&~\text{late time}~t\gg t_{\rm sd}\\
    &B_{12}^2P_{0,-3}^{-4}&\text{for}&~\text{early time}~t\ll t_{\rm sd}\\
\end{aligned}
\right.\nonumber\\\label{spow} \ee where $L_{\rm
    sd,0}=I\Omega_0^2/2t_{\rm
    sd}\simeq10^{47}~\unit{erg~s^{-1}}B_{14}^{2}P_{0,-3}^{-4}$ is the
initial spindown power, $t_{\rm
    sd}=3c^3I/B^2R^6\Omega_0^2=2\times10^5~\unit{s}B_{14}^{-2}P_{0,-3}^2$
is the spindown timescale, $I=10^{45}~\unit{g~cm^2}$ is the moment
of inertia of the neutron star, $\Omega_0=2\pi/P_0$ is the initial
angular velocity of the neutron star. According to the above
equation, in order to make the spindown luminosity order at the same
order of magnitude as the isotropic luminosity of FRBs when an FRB
occurs, for the late-time case, the neutron star is required to be a
magnetar with magnetic field $B\sim10^{14}~{\rm G}$, and for the
early-time case, the neutron star is required to be a normal pulsar
with magnetic field $B\sim10^2~{\rm G}$. Due to the rotation energy
loss, the spin period will increases with time, e.g., \be
P=P_0\left(1+\frac{t}{t_{\rm sd,0}}\right)^{1/2}\simeq1~{\rm
ms}\left\{
\begin{aligned}
    &12B_{14}t_{\rm yr}^{1/2}&\text{for}&~\text{late time}~t\gg t_{\rm sd}\\
    &P_{0,-3}&\text{for}&~\text{early time}~t\ll t_{\rm sd}\\
\end{aligned}
\right.\nonumber\\
\ee Therefore, for a magnetar with $B\sim10^{14}~{\rm G}$ and
$P_0\sim1~{\rm ms}$, when its period increases to $P\lesssim0.1~{\rm
    s}$, the magnetar age is required to be $t\lesssim100~{\rm yr}$. On
the other hand, if the radio bursts are powered by the
rotational energy, e.g., $L_{\rm sd}\gtrsim f_bL_{\rm FRB}$, where
$L_{\rm FRB}\sim10^{42}~{\rm erg~s^{-1}}$ is the FRB isotropic
luminosity, and $f_b$ is the beaming factor, the magnetar is
required to be very young with \be t<2~{\rm
    yr}f_b^{-1/2}L_{{\rm FRB},42}^{-1/2}B_{14}^{-1} \ee
The rotational energy seems not viable as the FRB power source for a
magnetar, the reason is as follows: 1. the spindown timescale of the
rapidly spinning magnetar would be shorter than the observation time
of FRB 121102 \citep{kat16a,kat18a}, which is of the order of
several years; 2. a young ejecta associated with a magnetar with age
of $t\ll1~{\rm yr}$ would involve an observable DM decreasing
\citep[e.g.][see next section]{pir16,yan17a}, which is against with
the observation of FRB 121102 \citep{hes19,jos19}; 3. the
distribution of DMs of non-repeating FRBs is inconsistent with that
of expanding SNR \citep{Katz16}. Therefore, the other energy powers
would be necessary, e.g., magnetic power \citep{Metzger17},
gravitational power \citep{gen15,dai16}, and kinetic power
\citep{zha17b,zb2018}.

\subsection{Observation properties from the ejecta of binary neutron
star mergers}

In this section, we consider the observation properties of the
ejecta of binary neutron star mergers. Different from supernova from
core-collapse explosion, the ejecta by BNS merger has higher
velocity $v\sim(0.1-0.3)c$ and lower mass
$M\sim(10^{-4}-10^{-2})M_\odot$. At the time $t$ after the BNS
merger, the ejecta electron density is \be n_e\simeq\frac{\eta Y_e
M}{4\pi m_pv^3t^3}=2.8~{\rm cm^{-3}}\eta Y_{e,0.2}
M_{-3}v_{0.2}^{-3}t_{\rm yr}^{-3}, \ee where $n_e$ is the free
electron density, $\Delta R\sim vt$ is the ejecta thickness,
$Y_e=0.2 Y_{e,0.2}$ is the electron fraction, $\eta$ is the
ionization fraction, $M_{-3}=M/10^{-3}M_\odot$, and
$v_{0.2}=v/0.2c$. Due to $\nu_{\rm FRB}\sim1~{\rm
    GHz}\gg\nu_p\simeq10^4~{\rm Hz}\sqrt{n_e}\sim17~{\rm kHz}$, the
ejecta plasma is transparent for FRB. Besides, for an extremely
young ejecta, the FRB emission may be subject to a large free-free
opacity, so that the FRB may not be detected. The free-free optical
depth is \be \tau_{\rm ff}=\alpha_{\rm ff}\Delta R\simeq(0.018T_{\rm
eje}^{-3/2}Z^2n_en_i\nu^{-2}\bar g_{\rm ff})\Delta
R=2.7\times10^{-8}\eta^2Y_{e,0.2}^2M_{-3}^2T_{\rm
eje,4}^{-3/2}\nu_9^{-2}v_{0.2}^{-5}t_{\rm yr}^{-5}, \ee where
$T_{\rm eje}=10^4T_{\rm eje,4}~{\rm K}$ is the ejecta temperature,
$\bar g_{\rm ff}\sim1$ is the Gaunt factor, $n_i$ and $n_e$ are the
number densities of ions and electrons, respectively, and $n_i\sim
n_e$ and $Z\sim 1$ are assumed in the ejecta. Thus, the ejecta will
transparent for the free-free absorption a few weeks after the BNS
merger.

Next, we discuss the dispersion measure and rotation measure from
the ejecta. The disperse measure contributed by the ejecta is \be
{\rm DM}=n_e\Delta R\simeq\frac{\eta Y_eM}{4\pi
    m_pv^2t^2}\simeq0.17~{\rm pc~cm^{-3}}\eta Y_{e,0.2}
M_{-3}v_{0.2}^{-2}t_{\rm yr}^{-2}. \ee We can see that the disperse
measure contributed by the ejecta is small. Only when
$t\lesssim1~{\rm yr}$, the ejecta can contribute an observable DM
variation. On the other hand, when the ejecta expand outward, the
magnetic flux in the ejecta would keep unchanged. The total
magnetic flux in the ejecta may be $\Phi\sim BR^2\simeq10^{26}~{\rm
G~cm^2}B_{14}R_6^2$. Due to the conservation of magnetic flux, one
has $\Phi\sim B_{\rm eje}(vt)^2\sim BR^2$, where $B_{\rm eje}$ is
the magnetic field strength in the ejecta. Finally, the RM
contributed by the ejecta at time $t$ is \be |{\rm RM}|=\frac{e^3
B_{\rm eje}}{2\pi m_e^2c^4}n_e\Delta R\sim\frac{e^3}{2\pi
m_e^2c^4}\xi BR^2v^{-2}t^{-2}{\rm DM}\sim4\times10^{-4}~{\rm
rad~m^{-2}}\xi\eta Y_{e,0.2} B_{14}R_6^2M_{-3}v_{0.2}^{-4}t_{\rm
yr}^{-4}, \ee where the magnetic configure factor $\xi$ is defined
as $\xi\equiv\left<B_\parallel\right>/\left<B\right>$. Notice that
RM would decrease faster than DM, since the magnetic field in the
ejecta also decreases with time, e.g., $B_{\rm eje}\propto t^{-2}$.
In summary, for the ejecta with age of $t\gtrsim1~{\rm yr}$, the
corresponding ${\rm DM}$ and ${\rm RM}$ are very small.

\section{Conclusions and discussion}
\label{sec:conclusions} Motivated by large offsets of FRB 180924, FRB 180916 and
FRB 190523, we study FRBs from activities of NSs newborn in
BNS mergers. FRB 180924 and FRB 190523 are localized to massive
galaxies with low star formation rate, which are dramatically
different to the host galaxy of FRB 121102. Firstly, we use the
latest binary-evolution code $\mathtt{BSE}$ to calculate properties
of NS-NS binaries. The merger time from population synthesis shows
similar distribution as gravitational wave delay time distribution,
i.e., $\ud N/\ud t \propto t^{-1}$ at late times. We show that the
host galaxies and offsets of FRB 180924, FRB 180916 and FRB 190523 are
well-matched to the distributions for NS-NS mergers from population
synthesis. In addition, using the galaxy model with similar mass as
short GRBs host galaxy, the offset distribution of short GRBs is
well reproduced from population synthesis.

The observational properties of FRBs from BNS merger channel are
also discussed. In this work, we consider that FRBs are formed by
coherent curvature radiation in the magnetosphere of neutron stars
\citep{Yang18}. Due to some accident events, e.g., neutron star
starquake \citep{wwy2018}, nearby astrophysical plasma
\citep{zha17b,zb2018}, or exotic small bodies \citep{gen15,dai16},
the magnetosphere is disturbed, and the fluctuation of net charges
with respect to the background Goldreich-Julian outflow would make a
coherent curvature radiation. The observed flux of FRBs requires
that the neutron star has large magnetic field $B\gtrsim10^{14}~{\rm
G}$ and fast rotation $P\lesssim0.1~{\rm s}$, which corresponds to a
young magnetar with age of $t\lesssim100~{\rm yr}$ and the initial
period of $P_0\sim1~{\rm ms}$. Meanwhile, since the ejecta of BNS
merger has high velocity $v\sim(0.1-0.3)c$ and low mass
$M\sim(10^{-4}-10^{-2})M_\odot$, the ejecta will be transparent for
free-free absorption a few weeks after the BNS merger, and the
corresponding DM and RM are very small for the ejecta with age of
$t\gtrsim1~{\rm yr}$.

% For an extremely young magnetar with age $t\lesssim1~{\rm yr}$, its spindown power could provide the FRB luminosity $L_{\rm FRB}\sim10^{42}~{\rm erg~s^{-1}}$. However, the rotation power of a magnetar is not viable as the FRB power, because
% a young nebula associated with a magnetar with age of $t\lesssim{\rm
%     a~few}~{\rm yr}$ would involve a large free-free opacity
% \citep{mur16,Metzger17,yan19} and an observable DM decreasing
% \citep{pir16,yan17a}. Besides, the spindown timescale of the rapidly
% spinning magnetar would be shorter than the observation time of FRB
% 121102, which is of the order of several years. Therefore, the other
% energy powers would be necessary, e.g., magnetic power
% \citep{Metzger17}, gravitational power \citep{gen15,dai16}, and kinetic
% power \citep{zha17b}.

% For a young magnetar formed from BNS merger,

In the BNS merger scenario, it would be possible that we observe
associations of FRBs with short GRBs and gravitational wave (GW)
events. In the future, if more FRBs are localized, the offset
distribution of short GRBs can be compared with that of FRBs. Since
the magnetar born in BNS merger can show magnetic activity for a
long time \citep{Beloborodov16,Beloborodov19}, we can search FRBs in the location
region of short GRBs and BNS merger GW events.

\section*{Acknowledgements}
We thank an anonymous referee for detailed and very constructive
suggestions that have allowed us to improve our manuscript. We thank
Sambaran Banerjee for sharing the updated BSE code. We also thank
Yong Shao and Xiangdong Li for helpful discussions. This work is
supported by the National Natural Science Foundation of China
(grants U1831207, 1573014, 11833003, 1185130, 11573021 and U1938104)
and the National Key Research and Development Program of China
(grant 2017YFA0402600).

\bibliography{ms}

%\begin{thebibliography}{dummy}
%\bibliographystyle{unsrt}

\begin{figure}
    \centering
    \includegraphics[width=0.6\textwidth]{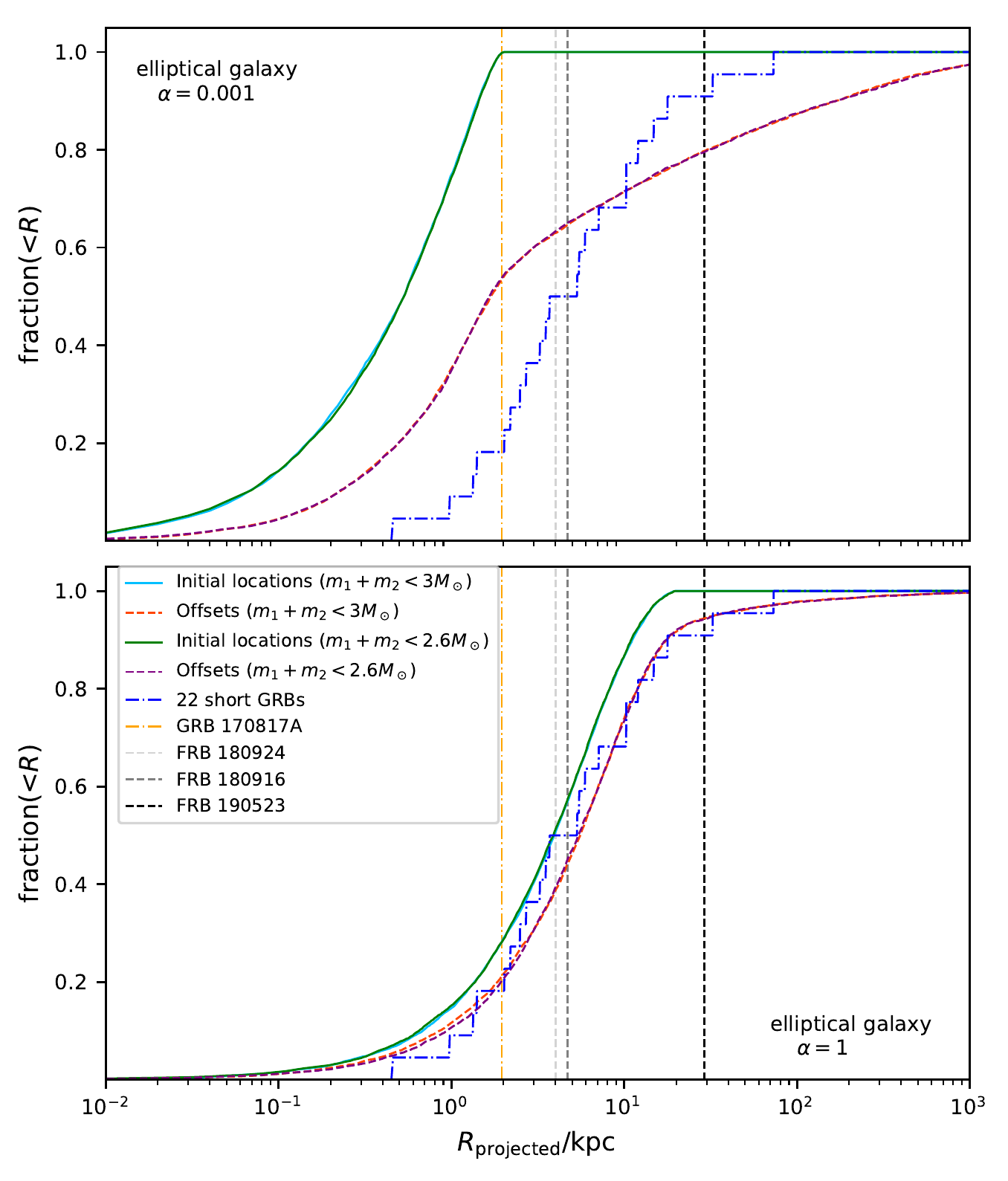}
    \caption{Initial location and offset cumulative distributions in elliptical galaxies for NS-NS mergers that may produce and observed offsets of short GRBs and FRBs.}
    \label{Offsets:BNS magnetar elliptical}
\end{figure}

\begin{figure}
    \centering
    \includegraphics[width=0.6\textwidth]{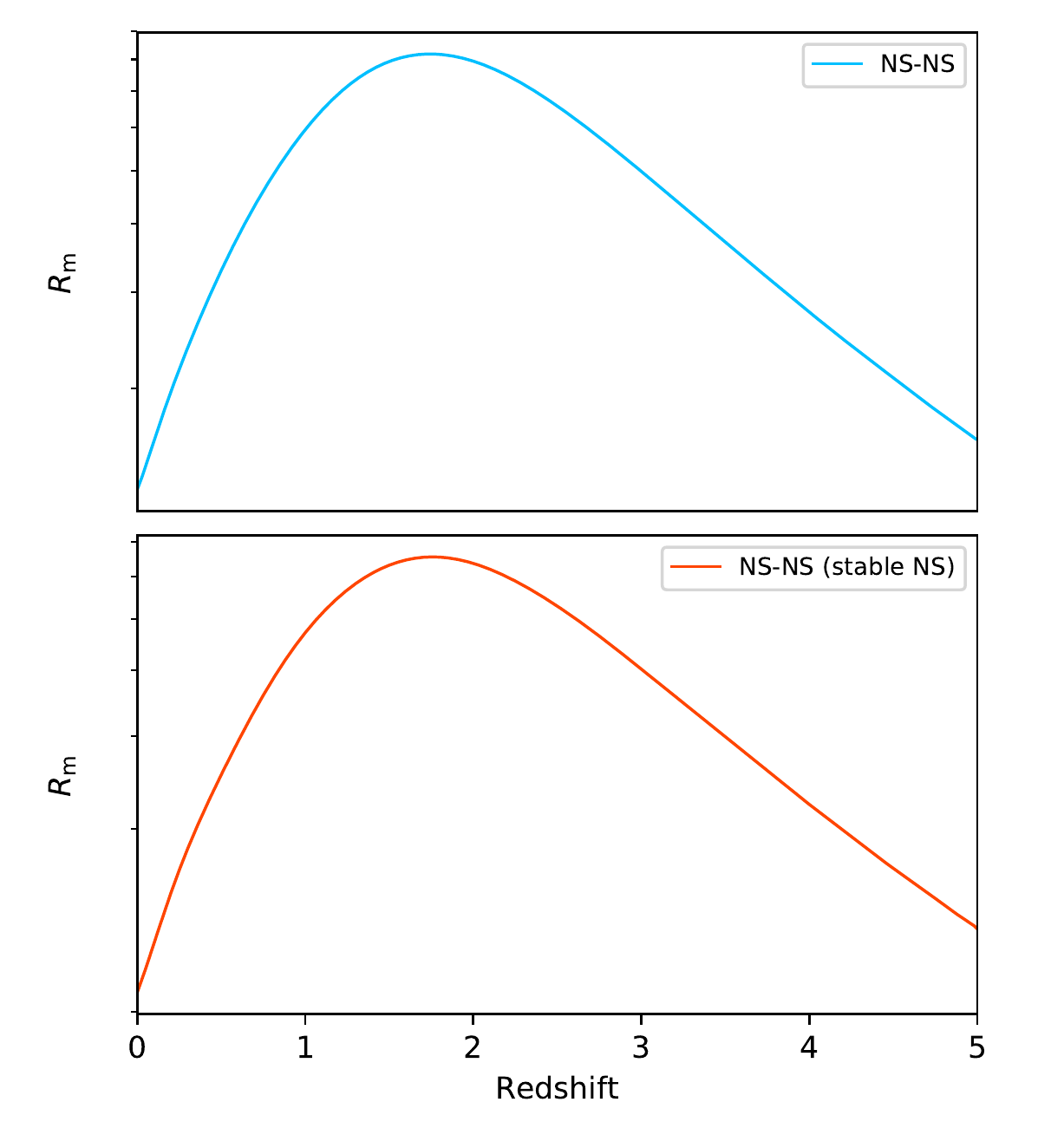}
    \caption{Merger rates $R_\mathrm{m}(z)$ as a function of redshift $z$ for NS-NS mergers (upper panel), and NS-NS mergers that can produce NSs (bottom panel). Rates are in arbitrary units.}
    \label{Merger rate: BNS}
\end{figure}

\begin{figure}
    \centering
    \includegraphics[angle=0,scale=0.6]{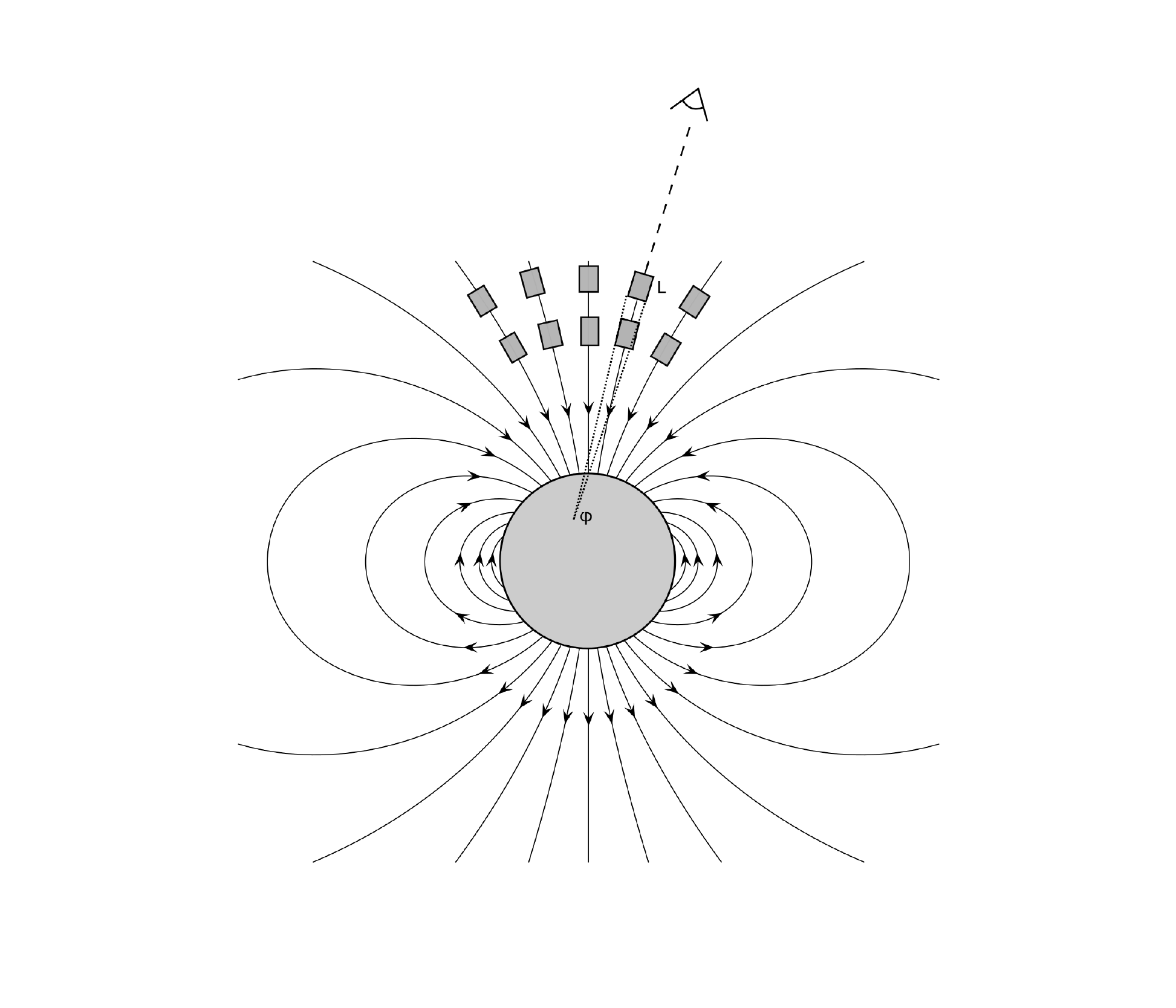}
    \caption{A cartoon figure for coherent curvature radiation by
        bunches in magnetosphere. The gray dark regions denote the bunches
        which are generated by the fluctuation of net charges with respect
        to the Goldreich-Julian background outflow. $L$ is the bunch length,
        and $\varphi$ is the bunch opening angle.}\label{fig1}
\end{figure}

\end{document}